\documentclass{egpubl}
\usepackage{pg2025s}

\WsConferencePaper

\usepackage[T1]{fontenc}
\usepackage{dfadobe}  
\usepackage{mathptmx} 
\usepackage{cite}  
\BibtexOrBiblatex
\electronicVersion
\PrintedOrElectronic
\ifpdf \usepackage[pdftex]{graphicx} \pdfcompresslevel=9
\else \usepackage[dvips]{graphicx} \fi

\usepackage{egweblnk} 
\usepackage{mathptmx}
\usepackage{times}
\usepackage{multirow}
\usepackage{colortbl} 
\usepackage{tabularx}
\usepackage{enumitem}
\usepackage[switch]{lineno}
\usepackage{amssymb}
\usepackage{tabu}                      
\usepackage{booktabs}                  
\usepackage{lipsum}                    
\usepackage{mwe}         
\usepackage{listings}
\usepackage{amsmath}
\usepackage{inconsolata} 
\usepackage[ruled,vlined,linesnumbered,noend]{algorithm2e}
\usepackage[table]{xcolor}%
\newcommand{\eg}{\emph{e.g.}}
\newcommand{\ie}{\emph{i.e.}}
\newcommand{\vs}{\emph{vs.} }
\definecolor{myred}{rgb}{0.7569, 0.1725, 0.1216}
\newcommand{\re}[1]{{\color{black}{#1}}}

\newcommand{\tool}{\emph{C2Views}\xspace}
\newcommand{\q}[1]{\textit{``#1''}}

\title[C2Views: Knowledge-based Colormap Design for Multiple-View Consistency]{C2Views: Knowledge-based Colormap Design for \\Multiple-View Consistency}

\author[Y. Hou et al.]
{\parbox{\textwidth}{\centering Yihan Hou$^{1}$\orcid{0000-0002-1459-8766},
        Yilin Ye$^{1,2}$\orcid{0000-0001-8874-5928},
        Liangwei Wang$^{1}$\orcid{0000-0003-3481-3993},
        Huamin Qu$^{2}$\orcid{0000-0002-3344-9694},
        and Wei Zeng$^{1,2}$\thanks{Wei Zeng is the corresponding author}\orcid{0000-0002-5600-8824}
        }
        \\
{\parbox{\textwidth}{\centering $^1$The Hong Kong University of Science and Technology (Guangzhou)\\
         $^2$The Hong Kong University of Science and Technology
       }
}
}
        
\begin{document}


\maketitle
\begin{abstract}
Multiple-view (MV) visualization provides a comprehensive and integrated perspective on complex data, establishing itself as an effective method for visual communication and exploratory data analysis.
While existing studies have predominantly focused on designing explicit visual linkages and coordinated interactions to facilitate the exploration of MV visualizations, these approaches often demand extra graphical and interactive effort, overlooking the potential of color as an effective channel for encoding data and relationships.
Addressing this oversight, we introduce \tool, a new framework for colormap design that implicitly shows the relation across views.
We begin by structuring the components and their relationships within MVs into a knowledge-based graph specification, wherein colormaps, data, and views are denoted as entities, and the interactions among them are illustrated as relations.
Building on this representation, we formulate the design criteria as an optimization problem and employ a genetic algorithm enhanced by Pareto optimality, generating colormaps that balance single-view effectiveness and multiple-view consistency.
Our approach is further complemented with an interactive interface for user-intended refinement.
We demonstrate the feasibility of \tool through various colormap design examples for MVs, underscoring its adaptability to diverse data relationships and view layouts.
Comparative user studies indicate that our method outperforms the existing approach in facilitating color distinction and enhancing multiple-view consistency, thereby simplifying data exploration processes.

\end{abstract}



\section{Introduction}
Multiple-view (MV) visualization is a popular method for effective visual communication and data analysis, often seen in dashboards\cite{bach2022dashboard} and visual analytics systems\cite{chen2020composition}.
Representing relations among views is essential to facilitate cross-view tasks and enhance visual analysis.
However, effectively depicting these relationships in multi-view settings can be challenging, particularly when dealing with complex data.
Current approaches rely on explicit visual linkages\cite{collins_2007_vislink, sun2021towards} or interactions\cite{weaver2009cross,chen_2021_nebula} to indicate coordination, potentially causing cognitive overload, highlighting the demand for implicit methods.

Implicit methods, such as view layout\cite{langner_2018_vistiles,wen_2023_effects} and data operations\cite{wu_2022_composition}, help maintain view consistency.
Color, a fundamental visual channel that operates independently of form and position, is well-suited for encoding data and indicating MV consistency simultaneously.
Effective colormaps enhance the effectiveness of single-view (SV) visualization and improve the consistency among multiple views.
However, designing a satisfactory colormap for SV visualization is challenging because it requires considering multiple criteria, such as color distinctness and harmony.
Designing colormaps for MV visualizations, which can have various types of data interrelations among views, adds complexity.

\begin{figure}[t]
    \centering
    \includegraphics[width=0.475\textwidth]{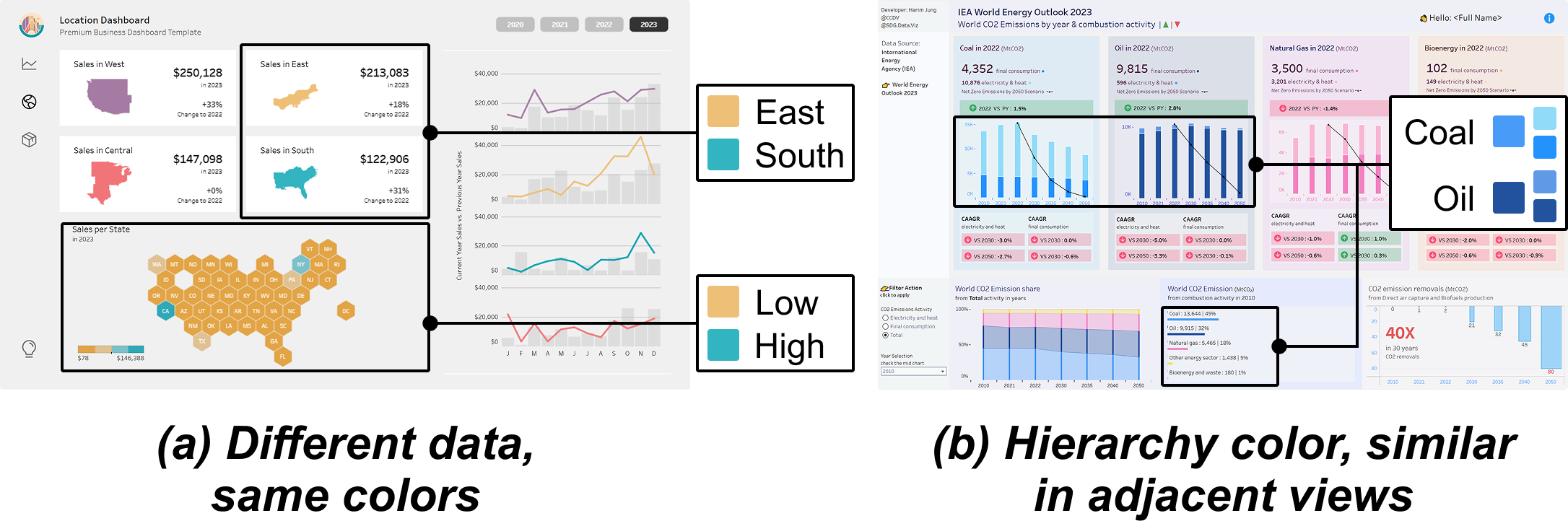}
    \caption{Common pitfalls for colormap design in MV visualization: a) different data with the same colormap, b) similar hieracy colormaps for adjacent views.}
    \vspace{-5mm}
    \label{fig:MVC_color_comparison}
\end{figure}

Existing authoring tools such as Tableau\cite{tableau} support iterative colormap design for individual views, but this manual process is challenging.
Novice users are often lack expertise in design principles for multi-view settings, leading them to unwittingly violate consistency rules\cite{qu2017keeping}.
\re{
    Even experts find balancing SV clarity and MV consistency arduous.}
\re{
    Common color design pitfalls appear even in well-crafted dashboards, such as those featured in Tableau's ``Viz of the Day'' community. 
    A frequent issue is ambiguity and confusion from inconsistent color usage (\eg, teal might represent different data in different views in Figure~\ref{fig:MVC_color_comparison}(a)). 
    A more complex issue arises when insufficiently distinct parent colors lead to problematic hierarchical palettes.
    In Figure~\ref{fig:MVC_color_comparison}(b), two shades of blue represent the parent categories ``Coal'' and ``Oil''. 
    While distinct individually, the sub-types' palettes overlap perceptually, hindering comparison across the two hierarchies.}

Automated colormap design can mitigate manual pitfalls, but current methods focus on SV visualizations, neglecting MV needs.
Our semi-automatic approach addresses this by balancing SV effectiveness with MV consistency, \re{tackling two challenges.
First, applying design principles for MV colormaps involves reconciling conflicting rules.
For instance, the need for discriminative colors for SV effectiveness\cite{lu2020palettailor, gramazio2016colorgorical} often clashes with the demand for uniform hues to represent hierarchical relationships across views.}
\re{Simply choosing one rule over the other leads to suboptimal design, making the balance of these competing demands a challenging trade-off that necessitates a multi-objective optimization approach to achieve a satisfactory solution.}
\re{Second, handling MV complexity requires a robust way to model intricate data-view relationships, influenced by factors like data and layout.}
While existing colormap recommendation approaches for MV visualizations \cite{shi2022colorcook} focus on consistency among views, they overlook detailed inter-view data relationships.
\re{To unify these factors into a computable format, a structured graph representation is essential.}

To address these challenges, we categorize MV colormap design considerations into: 1) SV effectiveness, including local discriminability, and 2) MV consistency, including global discriminability, hue uniformity, continuity and spatial proximity (Sect.~\ref{ssec:design_req}).
We propose a knowledge-based graph representation that unifies the relevant design components, with colormap, data, and view represented as entities and relationships among them represented as relations (Sect.~\ref{sect:kg_construction}).
Based on this model, we define design criteria and formulate an optimization problem, solved via a genetic algorithm with Pareto optimality to generate colormaps across all views (Sect.~\ref{sect:optimization}).
The automatic approach is complemented with an interactive interface for user-intended refinements (Sect.~\ref{ssec:interface}).
We present case studies that demonstrate the feasibility and expressiveness of our approach in different scenarios (Sect.~\ref{ssec:case}).
We also conducted a user study, and the results showed that our method outperforms the existing colormap recommendation approach, particularly in accomplishing MV consistency (Sect.~\ref{ssec:UserStudy}).

In summary, our work makes the following contributions:
\begin{itemize}
    \item Identification and distillation of MV colormap design considerations, covering SV effectiveness and MV consistency.
    \item A knowledge-based graph representation unifying design components, including colormap, data, view, and complex design criteria from relations of the components.
    \item Development of \tool, a semi-automatic approach using a Pareto-enhanced genetic algorithm for colormap generation, complemented by an interactive refinement interface, validated through user studies and interviews.
    \re{The source code and all supplementary materials are publicly available at \url{https://github.com/Sunnary2604/C2Views}.}
        
\end{itemize}
\section{Related Work}
\label{sec:related}
\textbf{Multi-View Visualization}.
MV visualization is widely used in visual analytics for data analysis\cite{roberts_2007_state} and in dashboards for visual communication\cite{bach2022dashboard, yang2025dashboard}.
Designing effective MV visualizations requires consideration from various perspectives.
Qu and Hullman outlined encoding constraints, including data cardinality, available encoding resources, and chart layout, to achieve consistency across multiple views\cite{qu2017keeping}.
Additional factors identified in the visualization community include data relationships\cite{sun_2021_sightbi, yngve_2021_semantic}, view composition\cite{wu_2022_composition, chen2020composition}, interaction coordination\cite{chen_2021_nebula}, and display modality\cite{langner_2019_multiple, zeng_2023_semi-automatic}.
Maintaining view consistency is a primary goal in MV visualizations, which can be achieved through explicit visual linkages (\eg,\cite{collins_2007_vislink, sun2021towards, weaver2009cross}) or implicit designs such as layout and color\cite{swayne1998xgobi, weaver2004building, gresh2000weave}.
Explicit methods risk visual clutter and interaction overload, while implicit designs leverage existing visual elements to avoid these issues.
For example, Wen et al.\cite{wen_2023_effects} used proper layouts to facilitate immersive analytics involving multiple views. 
In our work, we adopt an implicit approach, focusing on colormap design to enhance multi-view consistency and using the large design space of colors to align with complex data relationships.

\smallskip
\noindent
\textbf{Colormap Design}.
Color is a key visual channel for information perception and has attracted attention in visual perception\cite{holzschue_2011_understanding, maule2022development} and graphic design\cite{donovan_2011_color, guo_2022_creative}. 
In visualization, color is frequently used to encode data and facilitate user understanding\cite{silva2011using, zhou2015survey}.
Colormaps define the mapping from data values to colors, and extensive studies have examined perception and its implications for colormap design.
To ensure effectiveness, proposed guidelines include avoiding rainbow colormaps\cite{liu_2018_somewhere, reda_2021_rainbows}, creating data-dependent designs\cite{harrower2003colorbrewer}, optimizing perceptual discriminability\cite{lee_2013_perceptually, fang2016categorical, lu2020palettailor} and contrast\cite{mittelstadt_2014_methods, mittelstadt_2015_efficient}, ensuring harmony\cite{phan2017color}, and following color semantics\cite{lin_2013_selecting, setlur2015linguistic}.

Despite these principles, applying them in practice is challenging, especially for novices.
As a result, many efforts have been made to simplify the colormap design process.
Approaches fall into three categories:
\emph{template-based} approaches (\eg, ColorBrewer\cite{harrower2003colorbrewer}, VSUP\cite{michael_2018_vsup}) offer pre-curated colormaps for users to choose;
\emph{rule-based} approaches (\eg, PRAVDAColor\cite{bergman1995rule}, Palettailor\cite{lu2020palettailor} and CCC-Tool\cite{nardini_2021_automatic}) identify specific colormap design rules and develop algorithms for automatic creation;
and \emph{data-driven} approaches extract color mappings from existing designs using techniques such as color legend detection\cite{poco_2018_extracting}, deep learning\cite{yuan_2021_deep}, or color-space structure analysis\cite{smart2019color}.
These approaches are incorporated into libraries and tools like d3.js\cite{Michael_2011_d3}, Tableau\cite{tableau}, and PowerBI\cite{power_bi}.
\re{Other research has also extended colormap design to dynamic contexts like animations and multi-scale models\cite{waldin2016chameleon}.}

However, most studies focus on single-view visualizations, whilst little research exists on colormap design for MVs.
Some early works focus on maintaining color coordination among MV visualizations\cite{swayne1998xgobi, weaver2004building, gresh2000weave}.
\re{The most relevant is ColorCook\cite{shi2022colorcook}, a rule-based colormap optimization for dashboards guided by effectiveness and expressiveness. 
However, it treats all views uniformly, generating a single parallel palette, and thus overlooks the specific relational structure between views and more complex data relationships like hierarchies.
}
In contrast, our work comprehensively summarizes data and layout relations among views and formulates these relations in a knowledge-based graph to enable automated MV colormap design.

\smallskip
\noindent
\textbf{Automated Visualization Design}.
Recent years have seen growing interest in automated visualization design, which falls into two main categories: data-driven and rule-based approaches.
Data-driven methods apply machine learning and generative models to recommend designs from large visualization datasets\cite{ye2024generative}, and have proven effective for dashboards\cite{wu2021learning, lin2022dminer} and colormaps\cite{smart2019color, yuan2021infocolorizer}.
However, they heavily rely on large and high-quality datasets, which are challenging to obtain for MV visualizations.
Datasets such as VIS30K\cite{chen_2021_vis30k} and other real-world datasets (\eg,\cite{chen2020composition}) contain MV examples but lack colormap labels.
In contrast, rule-based approaches formalize design guidelines into quantifiable constraints and solve them via optimization, such as Voyager\cite{wongsuphasawat2015voyager} and Draco\cite{moritz2018formalizing}.
Nonetheless, defining appropriate rules for automated visualization design is labor-intensive, especially for MV visualizations.
For instance, DMiner\cite{lin2022dminer} mined over 600 rules from existing MV designs.
As the complexity of rules increases, the number of design constraints that must be satisfied also increases significantly, making optimization more challenging.

To manage the complexity of rule-based systems, we employ a knowledge-based graph representation. 
While this approach has been successfully used to model intricate rules in general visualization design\cite{gilson_2008_from,chen_2019_ontological, li2021kg4vis}, it has not been applied to colormap design specifically.
We take advantage of the representation of relations among colormap, data, and view entities to optimize colormaps for MV.
\section{Scope and Design Considerations}
\label{sec:over}

This section clarifies the scope of the work (Sect.~\ref{ssec:scope}), followed by design requirements and considerations (Sect.~\ref{ssec:design_req}).
\subsection{Method Scope}\label{ssec:scope}

Effective color usage is essential for conveying information and context in common MV applications like dashboards and visual analytics systems.
We conducted systematic reviews on colormap design in visualization\cite{silva2011using, zhou2015survey} and identified various design perspectives that need to be considered, including:
\begin{itemize}
      \item \textit{Color consistency in MV visualization}.
            Our primary focus is on MV visualizations, with a particular emphasis on color consistency across views based on data relationships.
            Nevertheless, we also recognize the importance of SV effectiveness, including considerations to ensure data correspondence and maintain graphical integrity in SV visualizations.
      \item \textit{Data mapping}.
            Besides mapping data values, color can also be a powerful tool for conveying meanings and semantics\cite{lin_2013_selecting,schloss_2018_mapping,yngve_2021_semantic,shi2022colorcook}.
            However, it is important to be aware of potential limitations, such as dissimilar background colors can significantly impact color perception and hinder effective data understanding \cite{mittelstadt_2015_methdos}.
            This work focuses on color mapping for data values rather than on reflecting meaning or semantics.
      \item \textit{1D colormap}.
            The focus of this study is on designing 1D colormaps for individual views rather than creating bi- or multi-variate colormaps for multiple views simultaneously.
            While there are scenarios in which bi-variate data must be presented together (\eg, \cite{zeng_2021_revisiting}), a variety of 2D colormaps (\eg, VSUP \cite{michael_2018_vsup}) have been developed to meet this need.
            However, these 2D colormaps have been criticized for being challenging to interpret, particularly when there are multiple levels in both attributes\cite{bernard_2015_survey}.
\end{itemize}

\begin{figure}[t]
      \centering
      \includegraphics[width=0.8\linewidth]{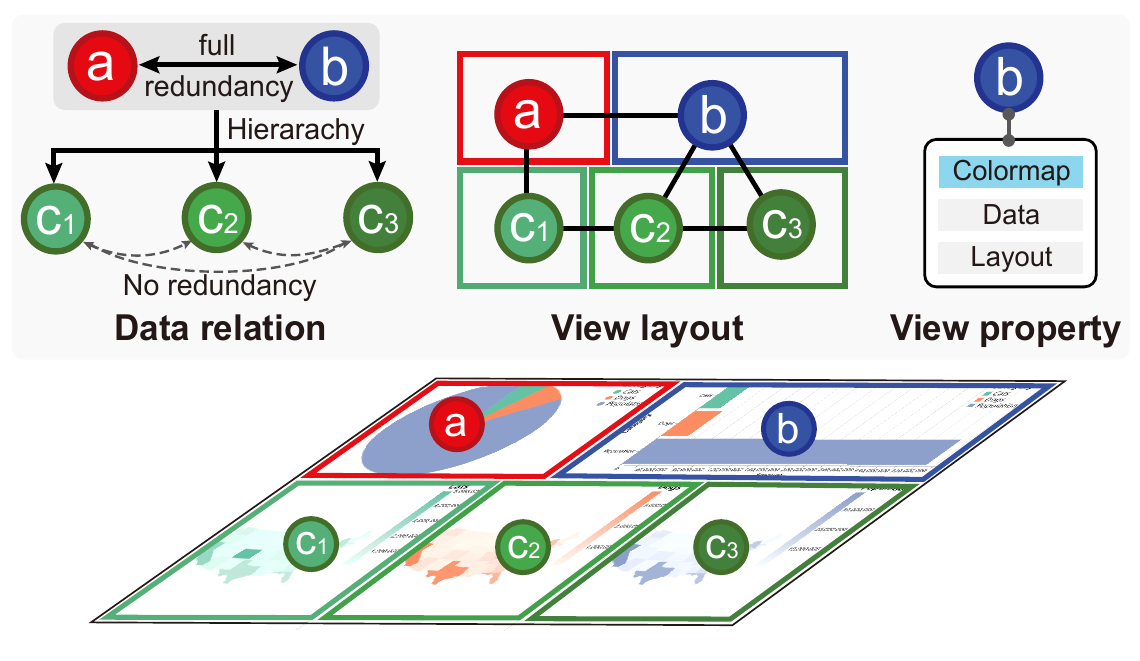}
      \caption{The main considerations for colormap design in MV visualizations in this work include \textit{data relation} and \emph{view layout}.}
      \vspace{-5mm}
      \label{fig:design_space}
\end{figure}


\begin{figure*}[t]
      \centering
      \includegraphics[width=0.95\textwidth]{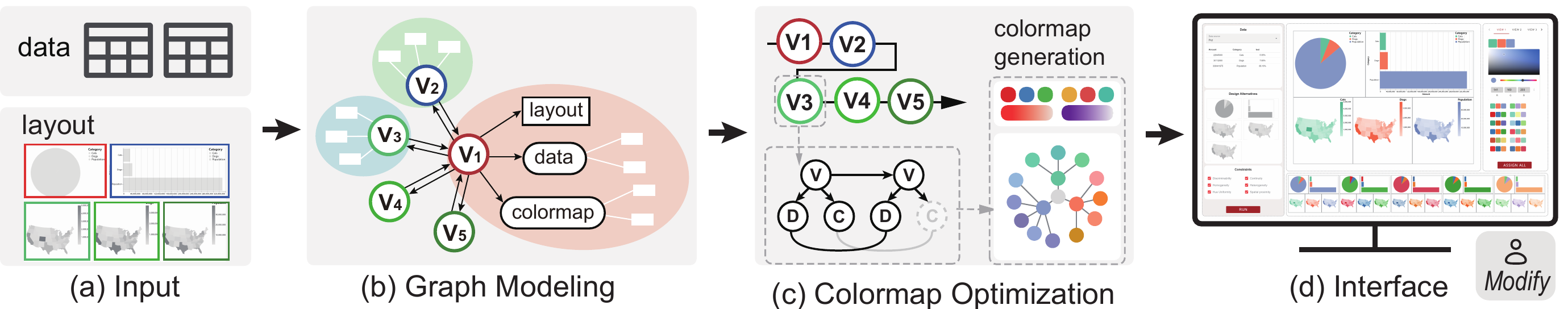}
      \caption{The proposed two-stage framework for colormap design on MV visualizations. (a) Input from the user, (b) knowledge-based graph construction representing the relationship between colormap, data, and view, (c) A knowledge-based graph optimization process for deriving an optimal colormap for multiple views, and (d) Interactive User Interface for customizing and adjusting the colormap used in each view.}
      \vspace{-5mm}
      \label{fig:framework}
\end{figure*}

\subsection{Design Requirements and Considerations}\label{ssec:design_req}
The manual design process often begins with selecting a colormap based on foundational rules like perceptual distance and effectiveness. 
However, for multi-view visualizations, this is usually followed by a complex series of manual adjustments, including assigning, checking, and tuning colors to ensure cross-view consistency\cite{shi2022colorcook}.
The aim of this work is to automate this complex procedure, considering the following two main requirements:

\begin{enumerate}[leftmargin=*, label=\textbf{R.\arabic*}]
      \item \textbf{Single-View (SV) Effectiveness:}
            To comply with the effectiveness principle\cite{munzner_2014_visualization}, the choice of colormap should convey unambiguous information.
            Numerous design guidelines have been proposed for colormap design to ensure effectiveness and assist users in tasks like identifying values.
            Although this work is intended for MV visualizations, achieving effectiveness in a single view is a main consideration.

      \item \textbf{Multiple-View (MV) Consistency:}
            MV visualizations provide flexibility for cross-view data analysis, such as comparing values in juxtapositioned views\cite{gleicher_2018_considerations}.
            Enhancing underlying relationships among views can assist users in gaining a comprehensive understanding of the data.
            This study aims to enhance view consistency using color, an implicit visual channel that requires no extra interaction or visual workload.
\end{enumerate}

\begin{table}[t]
      \caption{Design requirements and their corresponding criteria.}
      {\scriptsize
      \begin{tabular}{|c|m{2cm}|m{3.5cm}|}
            \hline
            \rowcolor[HTML]{EFEFEF}
            \textbf{Requirement}                                                            & \multicolumn{1}{c|}{\cellcolor[HTML]{EFEFEF}\textbf{Criteria}} & \multicolumn{1}{c|}{\cellcolor[HTML]{EFEFEF}\textbf{Goals}}                       \\
            \hline
            \begin{tabular}[c]{@{}c@{}}R1: SV \\ Effectiveness\end{tabular}                 & C1: Local Discriminability                                     & Identify marks\cite{gramazio2016colorgorical, lu2020palettailor}                  \\
            \hline
                                                                                            & C2: Global Discriminability                                    & Keeping consistence\cite{qu2017keeping, yngve_2021_semantic}                      \\
            \cline{2-3}
                                                                                            & C3: Hue Uniformity                                             & Enhancing hierarchy relation\cite{tennekes2014tree}                               \\
            \cline{2-3}
                                                                                            & C4: Continuity                                                 & Equal presence of data\cite{chuang2009hue, shi2022colorcook, kim2014perceptually} \\
            \cline{2-3}
            \multirow{-4}{*}{\begin{tabular}[c]{@{}c@{}}R2: MV \\ Consistency\end{tabular}} & C5: Spatial Proximity                                          & Balance constraints\cite{qu2017keeping}                                           \\
            \hline
      \end{tabular}
      }
      \vspace{-5mm}
\end{table}

To translate our design requirements into a computable model, we first consider the core components of the design space, including colormap, data, and view components, along with their relationships (Figure \ref{fig:design_space}).
\re{Based on this structure, we formulate our high-level requirements into the following five design criteria, which are then quantified as mathematical metrics in Sect.~\ref{sssec:metric}}.

Many studies have focused on enhancing the effectiveness of color encoding for SV visualization, which mainly focuses on color discriminability (\emph{C1}).

\begin{enumerate}[leftmargin=*, label=\textbf{C\arabic*.}]
      \item \textbf{Local Discriminability:}
            The ability to quickly and accurately identify a graphical mark and perceive differences among marks is a critical design criterion when representing categorical data\cite{gramazio2016colorgorical, lu2020palettailor}.
            Though influenced by other factors like name difference, color discriminability is primarily determined by the perceptual distance between color pairs in a uniform color space.
\end{enumerate}

In contrast, there are fewer guidelines available on designing colormaps for MV consistency, particularly use colormaps to represent cross-view relationships in MV.
To address this gap, we reviewed the relevant literature on MV design and identified key considerations related to \emph{data relationship}, including full/partial/no redundancy and hierarchy relations, as well as \emph{view layout} for the spatial distance between views.
We then related these considerations to colormap design and distilled a set of desirable criteria as follows:

\begin{enumerate}[leftmargin=*, label=\textbf{C\arabic*.}, resume]
      \item \textbf{Global Discriminability:}
            The consistency constraint model advocates for the use of consistent encoding for identical data and distinct encoding for different data\cite{qu2017keeping, yngve_2021_semantic}.
            Specifically, we use identical colors for the full/partial redundancy relations data.
            Sticking to the consistency model, data attributes that have no redundancy relations should be assigned distinct colors.
            The criteria not only considers color discriminability within a single view but also requires that the colors be distinguishable from each other in multiple views.
            To achieve this, the same metric for a perceptual distance of color discriminability can be used.

      \item \textbf{Hue Uniformity:}
            Hierarchical relationships across views are prevalent, connecting visual elements in one view to several in another\cite{sun_2021_sightbi}. To illustrate these connections, especially for categorical elements, the tree colormap can be utilized\cite{tennekes2014tree}.
            Effective tree colormaps deploy perceptually uniform hues to clarify hierarchical links among different elements.
            On the other hand, sequential colormaps with varying lightness levels but the same hue are recommended for ordinal attributes of the child data.
      \item \textbf{Continuity:}
            The Gestalt laws of continuity suggest that consistent use of colors can improve user preference and create a sense of unity\cite{wertheimer1938laws, kim2014perceptually, shi2022colorcook}.
            We prioritize the effective, even representation of attributes, avoiding the distortion caused by overly bright colors.
            Balancing color continuity and color discrimination can often be a challenge.
            However, we strive to achieve a balance between the two by adhering to the principles set forth in Colorgorical\cite{gramazio2016colorgorical}.
      \item \textbf{Spatial Proximity:}
            Meeting all of the above criteria within a limited color space can be challenging.
            Studies indicate that the layout of multiple views (MVs) can affect the constraints among them\cite{qu2017keeping}.
            For instance, global distinctness may not be as crucial when two views are not placed side by side.
            To address this, we calculate the spatial proximity between views and use it to assign weight to the aforementioned criteria.

\end{enumerate}

\section{\tool}
This section outlines the two-stage framework for achieving color consilience in an MV visualization, including \emph{Graph Modeling} (Sect.~\ref{sect:kg_construction}) that formulates the design considerations and related components in a graph, and \emph{Colormap Optimization} (Sect.~\ref{sect:optimization}) that optimize colormaps to balance SV effectiveness and MV consistency.
Moreover, we offer an \emph{Interactive Interface} (Sect.~\ref{sect:interface}) that enables user-driven refinement based on their preference.

\begin{figure}[t]
      \centering
      \includegraphics[width=0.99\linewidth]{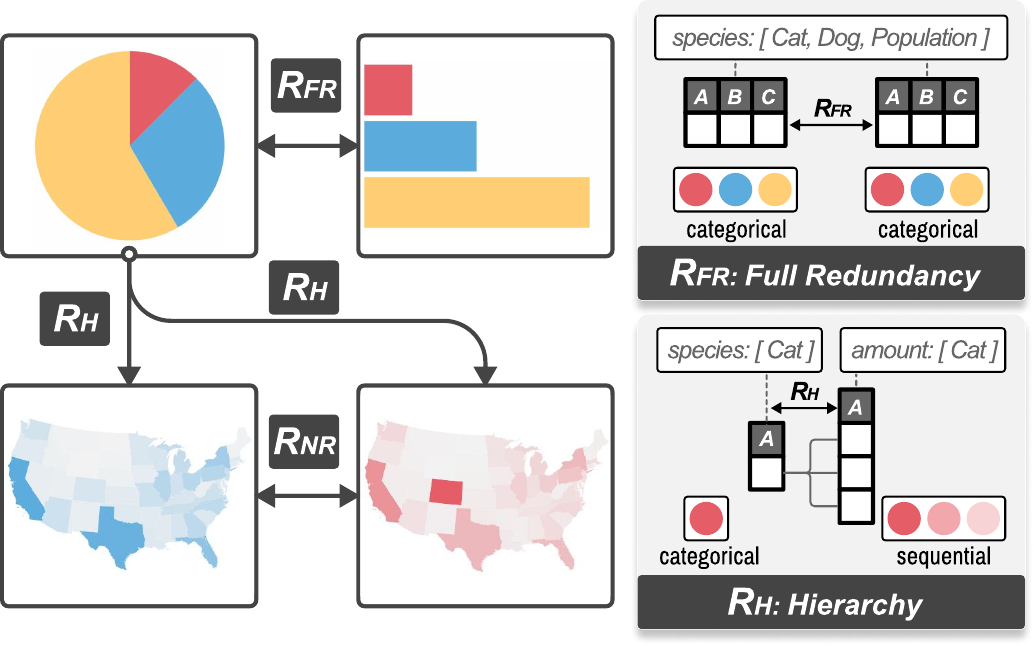}
      \caption{Example of data-data relations for color palette generation, including full redundancy and hierarchy.}
      \label{fig:data_relation}
      \vspace{-5mm}
\end{figure}

\subsection{Graph Modeling}
\label{sect:kg_construction}

To model the complex components and relationships for automated multi-view colormap design, we propose a holistic knowledge-based graph.
\re{This graph encodes instance-specific facts, such as data relationships and view layouts, inferred from user inputs to build a model for optimization. 
      The ``knowledge" here refers to a priori design principles from visualization literature, which are applied to derive color relations and complete the graph structure, rather than being directly encoded. 
      This process yields a computable model tailored to each specific visualization instance.}

\smallskip
\noindent
\textbf{Definition of Entity.}
In the knowledge-based graph, three types of entities are defined as follows:

\begin{itemize}[leftmargin=*]
      \item
            \emph{View}.
            An MV visualization consists of a set of views, denoted as $\{V_i | i \in\{1,\cdots,n\} \}$, where $V_i$ represents an individual view and $n$ represents the view count.
            Each view $V_i$ includes properties of a bounding box and chart type predefined by users.
            The bounding boxes determine the layout relationship of the views.

      \item
            \emph{Data}. An MV visualization can represent various datasets or multiple perspectives of the same dataset\cite{roberts_2007_state}.
            For simplicity, we focus on the view level and refer to the data encoded by a single view as $D(V_i)$.
            The data $D(V_i)$ may consist of multiple fields, but at least one field must be encoded using colors.

      \item
            \emph{Colormap}.
            As previously mentioned, this work uses colors for encoding data without taking into account color semantic mapping.
            Thus, each view utilizes a colormap to represent a single field in the corresponding data, denoted as $C(V_i) \rightarrow D(V_i)$.
            A colormap $C(V_i)$ can either be a \emph{discrete} or \emph{continuous} type, where \emph{discrete} colormaps consist of a limited number of colors, while \emph{continuous} colormaps include a gradual progression of colors without abrupt changes.
            To maintain consistency, we model both discrete and continuous colormaps as an ordered list of $m$ colors, denoted as $C(V_i) := \{c_j\}_{j=1}^m$, for which we evenly sample $m$ colors from continuous colormaps.

\end{itemize}

\noindent
\textbf{Definition of Relations.}
Within a single view, the relationship between the view $V_i$, its corresponding data $D(V_i)$, and the colormap $C(V_i)$ is straightforward.
Our focus is on the cross-view relationships between views and data.
To model these relationships, we first form a topology graph by connecting all adjacent views.
This topology graph allows us to quantify the view-view relations $R(V_i, V_j)$ between two views $V_i$ and $V_j$ based on their distance.
In contrast, the data-data relations $R(D(V_i), D(V_j))$ between $D(V_i)$ and $D(V_j)$ can be more complex.
We conducted a literature review and identified several common types of data-data relations, including:

\begin{enumerate}[leftmargin=*, label=\textbf{DR.\arabic*}]
      \item\textbf{Full Redundancy}:
      Two data sets $D(V_i)$ and $D(V_j)$ are considered to have a \emph{full redundancy} relation if and only if their color-encoding data fields are exactly the same, \ie, $\forall d \in D(V_i) \rightarrow d \in D(V_j)$ and vice versa.
      For example, in Figure~\ref{fig:data_relation}, the pie chart and bar chart share the same data fields, which are `\emph{species}' and its variables [`cat', `dog', `population'].
      Therefore, these data sets require identical colormaps to maintain consistency.

      \item\textbf{Partial Redundancy}: Two data sets $D(V_i)$ and $D(V_j)$ are regarded as \emph{partial redundancy} relation if they have some overlapping but are not identical, \ie, $D(V_i) \bigcap D(V_j) \neq \varnothing$ and $D(V_i) \neq D(V_j)$.
      To meet the consistency requirement, the color encoding for a variable existing in both data sets should be identical in both views.
      On the other hand, to meet the \emph{global discriminability} requirement (\emph{C2}), the colors should be distinct.

      \item\textbf{Non-Redundancy}:
      Two data sets $D(V_i)$ and $D(V_j)$ are regarded as \emph{no redundancy} relation if they have no overlaps, \ie, $D_c(V_i) \bigcap D_c(V_j) = \varnothing$.
      For example, in the two maps shown in Figure~\ref{fig:data_relation}, colors are used to encode cat' and dog' separately, and there is no overlap in the color mapping between the two maps.
      In such cases, the colormaps shall be distinct from each other to meet the \emph{global discriminability} requirement (\emph{C2}).

      \item\textbf{Hierarchy}:
      The relation occurs when one data set, $D(V_i)$, is subordinate to or derived from another data set, $D(V_j)$.
      For example, in the bottom-left map shown in Figure~\ref{fig:data_relation}, colors are used to encode the number of `cat' variables in each state, which is only an element of the data in the pie chart and the bar chart.
      The \emph{hierarchy} relation is directional and should be considered when determining the ordering of colormap assignment (Sect.~\ref{sect:optimization}).
      The colormaps should also satisfy the \emph{hue uniformity} requirement (\emph{C5}) with perceptually uniform hues to indicate the hierarchy relation.
      For instance, map's red sequential colormap is consistent with the red color used in the pie chart and the bar chart.

\end{enumerate}

\noindent
\textbf{Graph Construction.}
With the definitions of entities and relations, the graph $G$ is \re{automatically constructed} for each input multi-view instance.
The process begins by iterating through all pairs of views ($V_i, V_j$) to \re{infer} their relationships based on the user's input. 
\re{
View-view relations are determined by calculating the spatial distance between their bounding boxes, adding an edge $E_{V} = (V_i, V_j)$ for adjacent views. 
Concurrently, the system infers data-data relations by analyzing the underlying data fields for patterns such as Full Redundancy (shared fields) or Hierarchy (aggregations). 
Once the relationship type $R_{ij}$ is identified, a corresponding typed edge $E_{D}=(D(V_i), R_{ij}, D(V_j))$ is added to the graph.} 
This iterative process results in a comprehensive graph that programmatically captures all necessary constraints for optimization. 

\subsection{Colormap Optimization}
\label{sect:optimization}
Optimization begins by translating our design considerations into a series of measurable metrics.

\subsubsection{Metric Formulation}
\label{sssec:metric}
We first leverage the following color metrics:

\begin{itemize}[leftmargin=*]
      \item
            \textbf{Color Difference.}
            To measure the color difference, which is crucial for local discriminability (\emph{C1}) in a single view and global discriminability (\emph{C2}) across multiple views, we use the CIEDE2000 color difference formula \cite{sharma2005ciede2000}.

            Here we measure the local discriminability of categorical colors $C(V_m)$ in a single view $V_m$ as:
            \vspace{-1.5mm}
            \begin{equation}
                  \small
                  f_{ldis}(V_m) =  \sum_{i=1}^{|C(V_m)|}\sum_{j=1,j \ne i}^{|C(V_m)|}{DE_{00}(c_i, c_j)}.
            \end{equation}
            A large value of $f_{ldis}(V_m)$ indicates that more distinct colors are used in $V_m$.
            Similarly, we compute the global discriminability of colors in two views $V_m$ and $V_n$ as:
            \vspace{-1.5mm}
            \begin{equation}
                  \small
                  f_{gdis}(V_m, V_n) =  \sum_{i=1}^{|C(V_m)|}\sum_{j=1}^{|C(V_n)|}{DE_{00}(c_i, c_j)}.
            \end{equation}
            A high value of $f_{gdis}(V_m, V_n)$ indicates that more distinct colors are used in $V_m$ and $V_n$.

      \item
            \textbf{Hue Uniformity.}
            To maximize the distance between different child views in the hierarchy (C3), we need to generate the parent view so that the color in the HSL space has the largest hue difference.
            The hue uniformity $f_{hu}$ is defined as below:
            \begin{equation}
                  \small
                  f_{hu}(V_m, V_n)=min(|H(c_i)-H(c_j)|), c_i \in V_m, c_j \in V_n,
            \end{equation}
            where $H(\cdot)$ stands for the hue value of a color.
            A larger value of $f_{hu}$ indicates that the two most similar primary colors of the subview are as far away from each other as possible, which is preferred.
      \item
            \textbf{Color Continuity.}
            To meet the requirement of color continuity across the whole MV (\emph{C4}), we measure the lightness difference in the CIELAB color space between colors in a colormap.
            \begin{equation}
                  \small
                  f_{con}(V_m, V_n) = \sum_{i=1}^{|C(V_m)|}\sum_{j=1}^{|C(V_n)|} (\left\vert L(c_i) - L(c_j) \right\vert),
            \end{equation}
            where $L(\cdot)$ stands for the lightness values of a color.
            A smaller value of $f_{con}$ indicates that more continuous colors are used.

\end{itemize}
\smallskip
\noindent
\textbf{Spatial Proximity.}
For criteria involving two views, namely $f_{gdis}$, $f_{hu}$, $f_{con}$ we weight the constraints based on view layout.
Specifically, we take into account the spatial proximity (\emph{C5}) between a pair of views $V_m$ and $V_n$ by incorporating the view layout into the weight function $\omega_{sp}$.
\vspace{-1.5mm}
\begin{equation}
      \small
      \omega_{sp} (V_m, V_n)=\frac{1}{D(V_m, V_n)},
\end{equation}
where $D(\cdot)$ is the distance between two views.
We organize all the views into a graph structure according to their adjacency relations and measure the spatial proximity between two views in terms of the shortest distance between two entities in the graph.
Smaller $\omega_{sp}$ indicates the spatial proximity between two views is weaker, and the corresponding constraints could be loosened.

\smallskip
\noindent
\textbf{Global Cost Formulation with Pareto Optimality.}
We normalize all the costs above to $[0 - 1]$ and define the global cost function for colormap for the whole MV.
The goal is to find a colormap $m$ that effectively balances the trade-offs between single view effectiveness ($C_{SV}$) and multiple view consistency ($C_{MV}$), each defined as the weighted sum of the individual costs as follows:
\vspace{-1.5mm}
\begin{equation}   
      \small
      C_{SV}(m)=\sum_{i=1}^{|V|}-\omega_{d}f_{ldis},
\end{equation}
\vspace{-2mm}
\begin{equation}   
      \small
      C_{MV}(m)=\sum_{i=1}^{|V|}\sum_{j=1, j \ne i}^{|V|}\omega_{sp}(-\omega_{gdis}f_{gdis}-\omega_{hu}f_{hu} + \omega_{con}f_{con}),
\end{equation}
where $\omega _{sp}$ applied to criteria involving two views.
The weights $\omega_{d}$, $\omega_{con}$, $\omega_{hu}$, and $\omega_{gdis}$ are all positive and reflect the relative importance of each factor; they are set to 1 by default.
We then employ a Pareto optimality approach to define the global cost function $G$.
A colormap $m$ is said to be Pareto optimal if there is no other colormap $m'$ such that it has a lower cost in both $C_{SV}$ and $C_{MV}$.
The Pareto front ${P}$ is the set of all such Pareto optimal colormaps and is defined as:
\vspace{-1.5mm}
\begin{equation}
      \small
      P = { m \in M \mid \not\exists m' \in M : (C_{SV}(m') < C_{SV}(m)) \land (C_{MV}(m') < C_{MV}(m)) },
\end{equation}
where ${M}$ is the set of all possible colormaps.
From this Pareto front, we select colormaps that best satisfy the design requirements and constraints, ensuring that the final colormap is both effective for single views and consistent across multiple views.




\subsubsection{Colormap Optimization}
We determine the coloring sequence according to the cross-view relationships in the graph, prioritizing views without dependencies such as hierarchical relations.
\re{We employ a genetic algorithm to efficiently explore the complex, multi-objective search space that results from generating colors in the continuous color space rather than from a discrete palette. 
A genetic algorithm is particularly suitable for this task due to its population-based search strategy and flexible encoding.
}
The algorithm takes the graph structure as input and iteratively optimizes the colormaps for all views in the multi-view visualization.
The following steps outline the procedure involved in this iterative process:

\smallskip
\noindent
\textbf{Population Initialization.}
The genetic algorithm commences by initializing a population of colormaps, each corresponding to the unique data fields for color encoding within the MV.
ColorBrewer and Tableau palettes, while not specifically crafted for MVs, offer a distinctive and broad coverage of the color space, ranging from 3 to 12 colors, making them good enough for initial seeding.
Predefined palettes are utilized directly when the required color count is within their limit.
When the necessary number exceeds what is available from predefined sets, we procedurally generate additional colors to supplement and enlarge the initial colormap selection.

\smallskip
\noindent
\textbf{Fitness Score Evaluation.}
All individuals in the population undergo a fitness score evaluation determined by a global cost function and Pareto optimization.
Colormaps that fail to meet hard constraints, such as insufficient color differentiation, are immediately rejected.
This process yields a Pareto front comprising a set of colormaps that optimally balance single-view effectiveness with multiple-view consistency.

\smallskip
\noindent
\textbf{Selection and Crossover.}
To preserve high-quality colormaps, we employ an elitist strategy where a set of top-performing individuals are selected from the Pareto front to form the elite pool for the next generation. 
We then perform crossover by randomly pairing colormaps from the Pareto front and swapping dominant colors with a 50\% chance.
This process passes on advantageous features while infusing the new generation with variation.

\begin{figure}[t]
      \centering
      \includegraphics[width=0.95\linewidth]{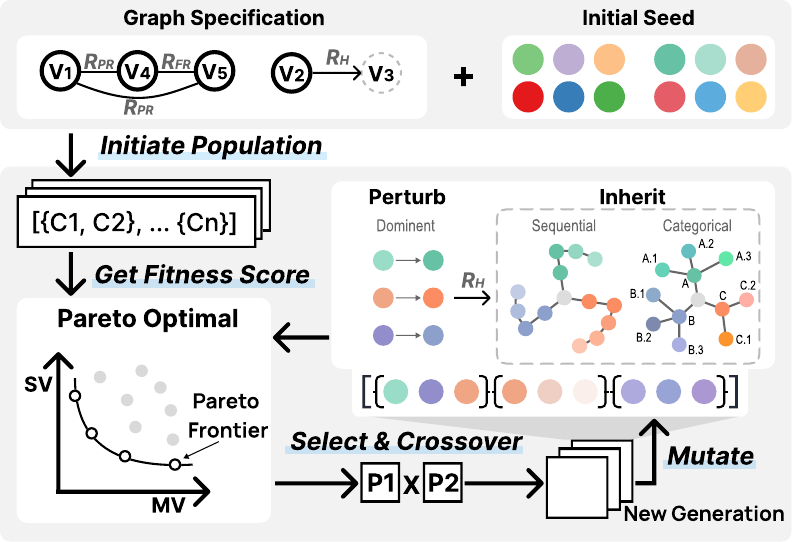}
      \caption{Optimizing colormaps via a Pareto-enhanced genetic algorithm. Inputs include multi-view specifications and an initial color palette. The iterative process yields colors that balance SV effectiveness with MV consistency through Pareto optimization.}
      \label{fig:hierarchy}
      \vspace{-5mm}
\end{figure}

\smallskip
\noindent
\textbf{Mutation.}
The genetic algorithm plays a pivotal role in the optimization process by ensuring variety within the population while preserving the beneficial traits from previous iterations.
This is achieved through two key operations: perturb and inheritance.
\begin{itemize} [leftmargin=*]
      \item \textit{Perturb}:
            This phase is fundamental for adding diversity to the population.
            By adjusting the dominant colors of the existing colormaps within a predetermined range of steps, we maintain a delicate balance.
            This controlled tweaking of the colormap's colors is essential for fostering diversity and retaining the successful attributes identified in earlier rounds.
      \item \textit{Inherit}:
            For hierarchical relations, the operation essentially involves generating a set of related colors inherited from a dominant color.
            Specifically, we adopt different approaches based on whether data in the child node are sequential or categorical or sequential.
            For sequential data, we modify the luminance and chroma in the HCL color space while keeping the hue consistent with the single color input to generate the corresponding sequential colormap.
            For categorical data, we draw inspiration from Tree Colors\cite{tennekes2014tree} for hierarchy-structured data, where the initial color is centered in the HCL color space, and the hue is adjusted within a certain range to generate colors with a similar hue.
            Figure \ref{fig:hierarchy}(a) illustrates a general case of this process, where the three marked colors  (A-green, B-blue, and C-orange) represent the initial colormap.
            We expand horizontally on the color ring to generate three categorical colormaps based on the hue of the parent color, \ie, A.1-3 inherited from A, B.1-3 inherited from B, and C.1-2 inherited from C.
            We reorder the colors in each generated colormap to ensure the distance between the two adjacent visual marks is uniformly perceived.
\end{itemize}

After executing multiple iterations of the genetic algorithm, a set of Pareto optimal colormaps emerges.
These colormaps are not only fine-tuned for individual view efficiency but are also consistent across the entire MV visualization, providing a superior user experience and enhanced data comprehension.

\subsection{Interactive User Interface}
\label{ssec:interface}

\begin{figure}[t]
      \centering
      \includegraphics[width=0.495\textwidth]{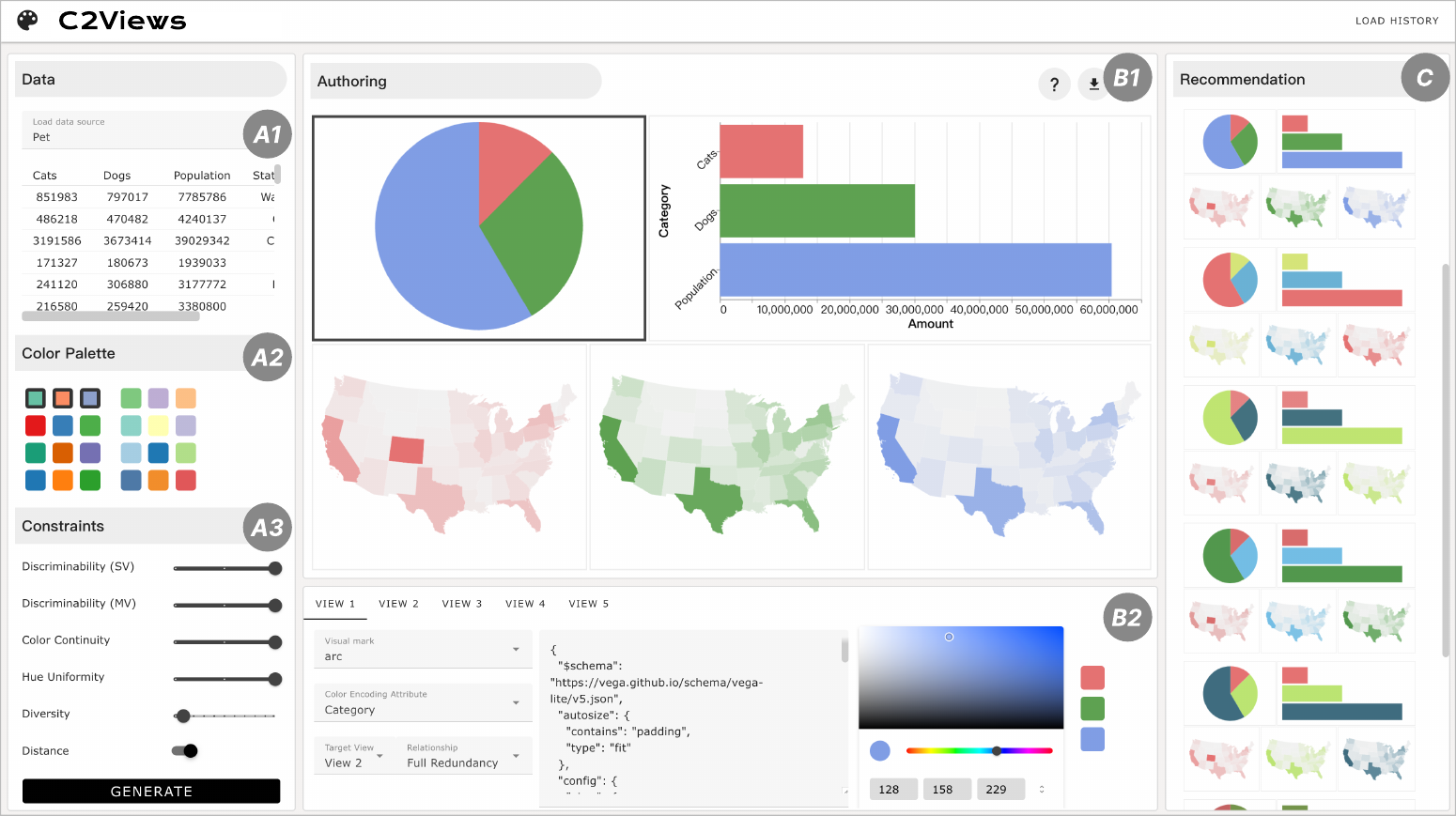}
      \caption{ System interface consists of a Control Panel (A), an Authoring Panel with editing detail (B), and a Gallery (C).}
      \vspace{-5mm}
      \label{fig:Interface}
\end{figure}

\re{
While our automated approach generates optimized colormaps, relying solely on automation may not be sufficient.
Manual refinement is essential to address subjective needs like aesthetics or branding, which are difficult to capture with perceptual metrics.
To support this, our interactive interface supplements automation with manual control, allowing users to fine-tune the resulting colormaps. 
}
As shown in Figure \ref{fig:Interface}, the interface has three components:
\label{sect:interface}

\begin{itemize}[leftmargin=*]
      \item  \textbf{Control Panel.}
            Users can load their dataset, view, select from predefined color palettes, and alter constraint settings.
            This panel presents a table view featuring the basic properties of the chosen dataset (Figure~\ref{fig:Interface}(A1)).
            Predefined color palettes are available in the color panel (Figure~\ref{fig:Interface}(A2)). 
            Users can tweak the weight of each constraint in the constraint panel (Figure~\ref{fig:Interface}(A3)).
            Once settings are satisfactory, users can generate colormaps for the target MV visualization by clicking on the ``GENERATE" button.
      \item \textbf{Authoring Panel.}
            This panel showing the MV visualization with the chosen colormap applied (Figure~\ref{fig:Interface}(B1)).
            \re{The editing panel (Figure~\ref{fig:Interface}(B2)) allows users to fine-tune the color settings for each view. 
            Using a tabbed interface, users can select a view to inspect its properties, which display the visual mark being encoded, the colormap type, and the automatically calculated data relationship. 
            For direct modification, the panel provides a color picker to adjust specific colors, as well as an editable code view of the Vega-Lite specification.}
      \item \textbf{Colormap Gallery.}
            \re{
            Upon generation, the gallery displays a set of Pareto-optimal colormap solutions (Figure \ref{fig:Interface} (C)). 
            Each represents a different trade-off and serves as a high-quality starting point for refinement.}
            A user can select any of these options to apply it to the main Authoring Panel for detailed adjustments.

\end{itemize}

\re{To preserve consistency during manual refinements, the system employs an automatic propagation mechanism. 
When a user modifies a color in the Authoring Panel, the system references the knowledge-based graph to instantly recalculate and update colors in all linked views based on their data relationship. 
This lightweight approach ensures consistency is preserved across the entire multi-view display without the need for a full, computationally expensive re-optimization.}

\section{Evaluation}

This section presents a comparative feasibility case study of our approach against ColorCook~\cite{shi2022colorcook} (Sect.\ref{ssec:case}).
Following this, we compare the results generated through our approach with those by ColorCook in a user study (Sect.\ref{ssec:UserStudy}).

\begin{figure*}[t]
      \centering
      \includegraphics[width=0.9\textwidth]{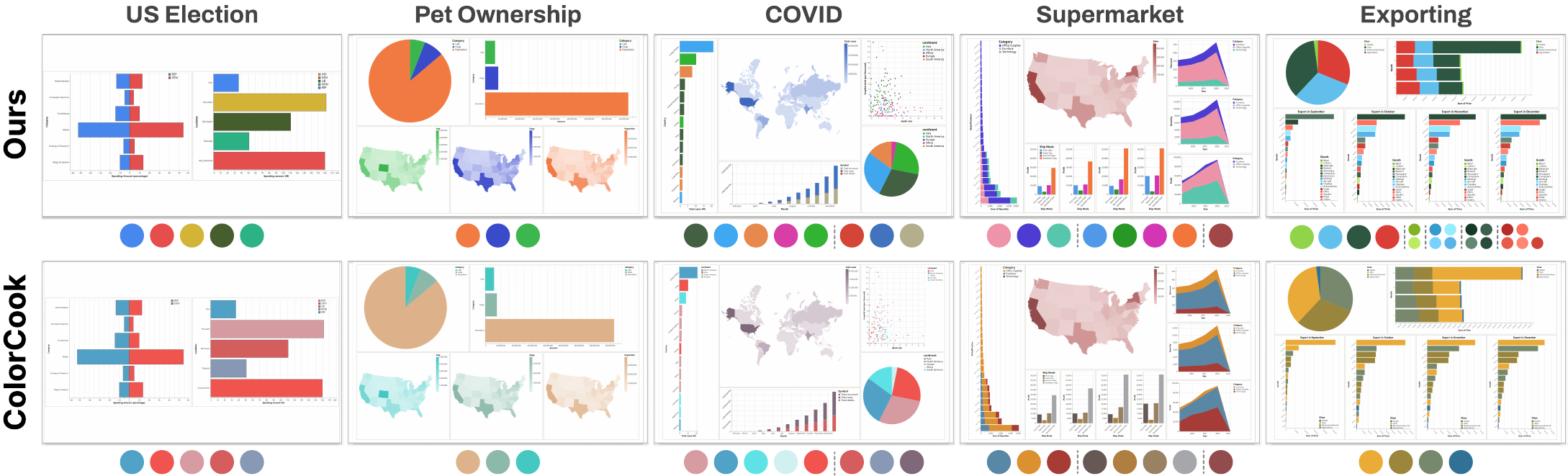}
      \caption{Examples of the result generated by ColorCook and \tool, with varying view counts, cross-view relationships, and data domains. The used color palettes are grouped and shown below. \re{More details are in the supplementary material.}}
      \vspace{-5mm}
      \label{fig:case}
\end{figure*}

\subsection{Case Study}\label{ssec:case}
Our method's feasibility and effectiveness are demonstrated across five cases with varying view counts, cross-view relationships, and data domains, as shown in Figure~\ref{fig:case}, benchmarking against the established ColorCook~\cite{shi2022colorcook} method.
For a fair comparison, all visualizations were generated automatically: we applied ColorCook's assignment algorithm using its most distinctive domain-appropriate palette, while our results were generated without manual refinement.

\smallskip
\noindent
\textbf{Partial Redundancy (Case 1)}.
This two-view visualization of US election data exhibits a partial redundancy relationship (Figure~\ref{fig:case}(A)).
ColorCook's generated less distinctive colormap for the right view, likely due to its reliance on specific dominant colors for domain semantics.
Furthermore, it is designed for a fixed number of categories (8 or 10), which limits flexibility for datasets with different category counts.
In contrast, our approach uses a broader hue spectrum to clearly differentiate the five parties across both views.

\smallskip
\noindent
\textbf{Hierachical (Case 2)}.
The second case introduces an MV visualization consisting of five views that provide an overview and detailed representation of pet ownership data, as shown in Figure~\ref{fig:case}(B).
As with the first case, ColorCook's color choices for the two sectors are less distinguishable with closing hue and low saturation, complicating sector comparison and the correlation of corresponding sectors between the top and bottom views.
Our method, however, uses a broader color spectrum with strong hue uniformity, improving the readability of the hierarchical relationship.

\smallskip
\noindent
\textbf{Use Multiple Group (Case 3 \& Case 4)}.
These cases demonstrate scalability of our method with more views and complex relationships. 
Case 3 presents an MV visualization comprising six views, divided into two color encoding groups.
The left bar chart, scatter plot, and pie chart use a 5-color palette to represent continent information with partial redundancy.
The bottom bar chart and map are color-coded to illustrate COVID symptoms, forming a hierarchical relation.
Increasing the number of views escalates the challenge of maintaining color distinctiveness, given the limited color space.
Unlike ColorCook, which treats all colorable sectors uniformly, our method considers both single-view and cross-view distinctiveness, facilitating a more effective color design.
Case 4 features a nine-view visualization with three color groups.
The left bar charts and the area charts employ colors to encode different categories of goods.
Four small multiples use color to represent shipping modes, while the map uses distinct colors for total sales.
ColorCook introduces confusion by using a similar red for unrelated data in the map and an area chart.
Our method avoids this by ensuring hue uniformity and color distinctiveness across all views, minimizing color overlap for clearer distinction.

\smallskip
\noindent
\textbf{Vertival hierarchical (Case 5)}
The final case presents a six-view MV visualization of export data, with a vertical hierarchical relationship: the top two views display total exports by main class, while the bottom four provide detailed breakdowns for each goods type.
Lacking explicit support for hierarchies, ColorCook incorrectly applies a full redundancy pattern, using the same colors for main classes at both levels.
In contrast, our method is designed to accentuate the hierarchical relationship between views, utilizing distinguishable colors for the top views and applying a tree colormap for the detailed bottom views.

One difference between the two methods is that ColorCook favors lower saturation for aesthetic reasons and employs similar hues for domain-specific palettes. 
In contrast, our approach emphasizes colormap effectiveness, resulting in higher saturation to enhance color distinctiveness.

\subsection{Quantitative Analysis}
\re{
      To objectively compare our method with the baseline, we evaluated the case studies using three metrics: Worst-Case Discriminability (WCD) for single-view clarity, the Parallel Relationship Score (PRS) for non-hierarchical consistency, and the Hierarchical Quality Score (HQS) for hierarchical encodings. For all metrics, a higher score indicates better performance, mathematically defined can be found in the supplementary material.
      
      The results, summarized in Table~\ref{tab:quantitative_comparison}, demonstrate a significant and consistent advantage for our approach. Our method achieved a WCD score of 32.84, substantially outperforming ColorCook's 11.61. A similar improvement was observed for parallel relationships, where our PRS score (27.64) was nearly three times higher than the baseline's (9.85).
      The most critical difference was in evaluating hierarchical structures. Our method achieved a positive HQS score of 1.50, whereas the baseline scored zero. This is because the baseline method does not account for hierarchical data, assigning the same color to all sub-categories and failing to visually represent the data's structure. In contrast, our positive score quantitatively demonstrates a successful encoding of these complex relationships.}

\begin{table}[ht]
      \centering
      \small 
      \caption{Quantitative comparison of our method against the baseline across three proposed metrics.}
      \label{tab:quantitative_comparison}
      \begin{tabular}{l c c c}
      \toprule
      \textbf{Method} & \textbf{WCD ($\uparrow$)} & \textbf{PRS ($\uparrow$)} & \textbf{HQS ($\uparrow$)} \\
      \midrule
      ColorCook       & 11.61 & 9.85  & 0.00  \\
      \textbf{Ours}   & \textbf{32.84} & \textbf{27.64} & \textbf{1.50}  \\
      \bottomrule
      \end{tabular}
      \vspace{-5mm}
\end{table}

\subsection{User Study}
\label{ssec:UserStudy}
We further conducted a user study to compare the results generated through our approach and those by the baseline methods of ColorCook~\cite{shi2022colorcook}.

\subsubsection{Study Design}
\noindent
\textbf{Data.}
We crafted a testing dataset comprising four diverse cases, each varying in complexity based on the number of views, the relation between views, and the underlying data domain.
All visualizations were generated automatically without manual refinement.
For the ColorCook component, we selected the most distinctive color palette appropriate for the data domain from their available dataset, aligning with their emphasis on color distinctiveness and aesthetic appeal to ensure a fair comparison.
Using these palettes, we then applied ColorCook's color-assignment algorithm to the MV visualizations.
To ensure participants concentrated on the color design, we minimized textual distractions by eliminating non-essential labels from the visualizations.
Essential text, including legends, was standardized across both methods in terms of font size to maintain readability.
In total, we have 8 distinct color designs for the 2 methods, with 4 cases each.

\smallskip
\noindent
\textbf{Participants.}
Our study enlisted 16 participants, balanced by gender, with ages ranging from 21 to 28 years (M=24.9, SD=1.87).
They came from diverse fields, including statistics, computer science, media art, and design, contributing a wide range of perspectives.
All participants reported experience in using or designing visualizations, with 9 participants who claimed to know little or had general knowledge, while 7 described themselves as familiar or very familiar with the subject.
In terms of color design, 10 had little or general knowledge, and 6 were familiar with the domain.
This diversity aimed to provide insights across different levels of expertise in visualization and color design.

\begin{figure}[t]
      \centering
      \includegraphics[width=0.9\linewidth]{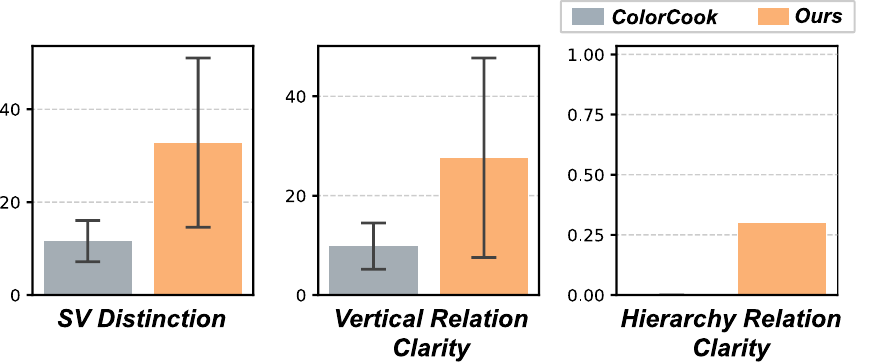}
      \caption{Quantitative results of the subjective ratings of different approaches on the cases.}
      \label{fig:quan_result}
      \vspace{-5mm}
\end{figure}

\smallskip
\noindent
\textbf{Procedure.}
The study was conducted in a quiet, controlled environment on a desktop computer, following four stages: introduction, experiment, questionnaire, and semi-structured interview. 
Participants first received a 5-minute overview of the research goals, key MV color design principles, and the role of color in illustrating cross-view relationships.
The evaluation metrics in Figure~\ref{fig:quan_result} were also introduced. 
After confirming their understanding, participants reviewed MV visualizations colored by both our method and ColorCook in a counterbalanced order, completing two single-view tasks (\ie, identifying maxima) and two cross-view tasks (\ie, multi-step inference using color cues).
To mitigate the learning effect, we slightly modified the tasks to avoid repetition, such as changing the retrieval range of values.
Afterward, participants filled out a questionnaire on a scale of 1-5 to evaluate the color designs based on the established metrics, with the designs anonymized as Design 1 and Design 2 to avoid bias.
The questionnaire consisted of 6 questions, with Questions 1-2 focusing on the ease of completing single-view and cross-view tasks, respectively.
Questions 3-5 assessed the colormap regarding color distinctiveness, relationship clarity, information conveyance, and preference.
Participants were encouraged to think aloud during the questionnaire phase.
After completing four cases, a semi-structured interview was conducted to gather qualitative feedback on the overall experience and practicality of the color designs.
The verbal responses were recorded and transcribed for further analysis.
Each session lasted about 40 minutes, and participants received \$10 compensation.

\subsubsection{Result and Analysis}
From the user study we obtained 640 responses (16 participants $\times$ 2 designs $\times$ 4 cases $\times$ 5 questions) from the user study.
We apply the Wilcox test to measure the significance level of the difference between the two methods, detailed in Figure~\ref{fig:quan_result}.
Significant values are reported for $p <.05(\ast)$, $p <.01(\ast\ast)$, $p <.001(\ast\ast\ast)$.
The results are summarized as follows:

\smallskip
\noindent
\textbf{Task Completion Ease (Q1 \& Q2).}
%
%
Our approach demonstrates a significant advantage over ColorCook in facilitating task completion (d.f. = 63, $p < .001$).
Specifically, our method (M=4.22, SD=0.65) outperforms ColorCook (M=3.25, SD=0.98) for single-view tasks, also demonstrating greater stability.
Cross-view tasks, which are inherently more complex due to the need for understanding relationships between different views, also see our method outperforming ColorCook (M=3.78, SD=0.98 vs. M=2.98, SD=1.07), underlining its enhanced efficacy in MV visualization exploration.

\smallskip
\noindent
\textbf{Color Distinctiveness (Q3).}
%
In the realm of color distinctiveness, our method (M=4.39, SD=0.76) significantly outperforms ColorCook (M=2.84, SD=1.08) (d.f. = 63, $p < .001$).
This is due to ColorCook's reliance on domain-specific and aesthetically pleasing colormaps, which may limit the range of available colors.
In contrast, our method emphasizes colormap effectiveness for discerning data attributes, resulting in clearer, more distinct color schemes.

\smallskip
\noindent
\textbf{Relationship Clarity (Q4).}
Our approach demonstrates enhanced clarity in visualizing view relationships (M=4.05, SD=0.98) compared to ColorCook (M=3.23, SD=1.10), with a significant difference (d.f. = 63, $p < .001$).
Such difference is mainly due to ColorCook using similar colors for non-related data attributes, which may lead to confusion.

\smallskip
\noindent
\textbf{Preference (Q5).}
There is no significant difference in user preference between our method (M=3.66, SD=0.96) and ColorCook (M=3.19, SD=1.22).
However, our method still maintains a slight advantage in user preference, indicating that our method is generally well-received by participants.

\subsubsection{Feedback}
In general, participants reported that using color to represent cross-view relationships implicitly was helpful, as it aided their understanding of the relationships between the data presented in each view and facilitated the completion of tasks.
Here are some highlights of the user feedback.

\smallskip
\noindent \textbf{\tool's Effectiveness.}
Most participants acknowledged the effectiveness of our approach in supporting task completion (N=14) since the generated colormaps have \q{higher construst} and \q{support identity different element}, making it effectively convey information (N=13).
As P5 noted, \q{The bright color allows me to associate information with the corresponding legend quickly.}
For multiple views, colormaps generated by \tool support quick linking between views, and \q{support finding the corresponding information at a glance} (P4, P15).
Despite the capability of accomplishing tasks using ColorCook colormap, it was reported to use colors of low saturation and similarity, which hinders efficient information seeking.
\q{It is hard for me to gain information in a single view, and I will be more likely to make mistakes in the cross-view.
      Using distinct colors will make me less nervous, and it will be easier to gain the information whether I am identifying differences or discovering relations}, P4 reported.

\smallskip
\noindent\textbf{Balancing Aesthetics with Information Efficiency.}
The majority of the participants prioritize the effectiveness and efficiency of information seeking and would like to use colormap generated by \tool in real practice (N=9).
They perceive it as a means to \q{minimize data interpretation efforts} and \q{swiftly access desired information}, deeming it \q{sufficient for data exploration}.
Despite being less effective, ColorCook is appreciated for its aesthetics and domain relevance (N=4), especially by designers and artists.
They note that they would modify the palette for improved distinction prior to real-world usage, as the existing colormap could impede task fulfillment.
An insightful observation from P8 highlighted that designers are particularly sensitive to subtle color differences, and ColorCook is distinct enough for them.
In such cases, they tend to use the harmonious one.
Similarly, P6 advocated for considering aesthetics when the discriminability is sufficient.
P2 added that the trade-off between aesthetics and effectiveness hinges on context: aesthetics take precedence in posters or presentations, whereas effectiveness is crucial for analysis.

\smallskip
\noindent\textbf{Navigating View Relationships with Color Guidance.}
Participants mentioned the generated color by \tool can help them get the connections between views quickly (N=6).
They can easily understand the full/partial redundancy, as well as the hierarchical color for sequential data.
However, they reported that the tree colormap should be carefully used.
Although half of the participants said they could understand the tree colormap, it still confused them at first glance and took time to understand (N=9).
They consider the color is \q{different} beyond the \q{hierarchical relationship} and prefer using consistent color across the hierarchical category data.
P14 noted that the effectiveness of the tree colormap hinges on the audience's expertise. It may be complex for the general public but offers simultaneous overview and detail to informed users, proving particularly beneficial for administrators.

\section{Discussion}

In this section, we reflect on our approach, demonstrate its possible application scenarios, and discuss the limitations and future work.

\begin{figure}[t]
      \centering
      \includegraphics[width=0.985\linewidth]{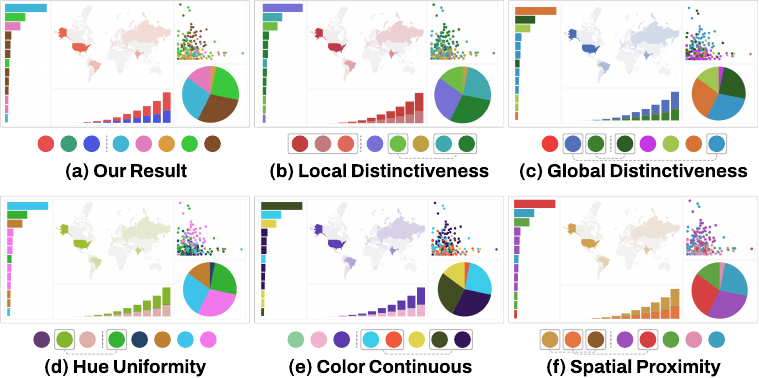}
      \caption{Example results of colormap generation with different metrics omitted from the cost function. (a) colormap considers all metrics. (b) - (f) are colormap results when omitting local distinctiveness, global distinctiveness, hue uniformity, continuity, and spatial proximity, respectively.}
      \vspace{-3mm}
      \label{fig:ablation}
\end{figure}

\smallskip
\noindent
\textbf{Contribution of Metrics.}
Our methodology incorporates multiple criteria, and assessing the impact of these metrics on the outcome is essential.
We investigate the effects of excluding specific metrics from the cost function:
\begin{itemize}[leftmargin=*]
      \item \textit{Local distinctiveness.}
            Exclusion of this metric results in color repetition within the same view, leading to a dominant color per view due to the requirement of global distinctiveness for color variation across views (Figure \ref{fig:ablation}(b)).
      \item \textit{Global distinctiveness.}
            Without this metric, colormaps may overlap between views (Figure \ref{fig:ablation}(c)), causing confusion despite well-distributed colors in each view.
            For example, views may inadvertently share a "green, blue" color scheme.
      \item \textit{Hue Uniformity.}
            Omitting hue uniformity in a sequential colormap causes monopolization of hues, leaving other views to utilize colors with the same hue but altered lightness and saturation.
            This can make it challenging to associate the colormap with its respective view (Figure \ref{fig:ablation}(d)).
      \item \textit{Continuity.}
            The absence of this metric leads to an imbalanced color distribution (Figure \ref{fig:ablation}(e)), making some colors disproportionately conspicuous and potentially distorting the intended focus, in agreement with Shi et al.~\cite{shi2022colorcook}.
      \item \textit{Spatial Proximity.}
            The aforementioned metrics can be conflicting.
            Discarding spatial proximity can lead to an inability to satisfy the established criteria and risks contravening both local and global distinctiveness (Figure \ref{fig:ablation}(f)).
\end{itemize}
The results confirm that all factors successfully constrain suboptimal outcomes, highlighting their roles in the optimization process.

\smallskip
\noindent
\textbf{Graph Representation and Optimization.}
The current optimization approach generates colormaps for all views simultaneously, which may neglect the complex relationship between views.
In contrast, our data relation-based optimization algorithm considers the relationship between views by utilizing a knowledge-based graph's (entity, relation, entity) representation.
This approach determines the color generation strategy based on the data relation and optimizes the global cost to generate colormaps that reflect the relationships between views.
We employ a genetic algorithm with Pareto optimality to optimize and balance the trade-offs between different metrics.
This is more effective than using a single weighted sum cost function, which may lead to suboptimal solutions.

\smallskip
\noindent
\textbf{Implicite \vs Explicit Encoding.}
The participants agreed that color is an important and useful channel to implicitly encode data relationships.
Particularly, they think the consistency of colormap serves as a visual cue for them to grab the key information without reading the text.
However, they also point out that many people are accustomed to combining text and color to reassure themselves of the reliability of perceived information.
\emph{``While I can infer the relationship between the views from the colors, I still need words to verify my hypothesis, and explicit representation can make me more confident"}, P7 commented.
This reflects that implicit representations are intuitive, but the combination of explicit and implicit representations is more desirable in practice.
Therefore, for the MV design, it is essential to balance implicit and explicit elements, conveying the information precisely and concisely.

\smallskip
\noindent
\textbf{Generalization and Application.}
The feedback from our user study highlights the importance of balancing explicit text and implicit color encoding.
Our implicit color encoding approach has potential benefits for various scenarios.
In common cases, our approach enables users to notice data relationships between views while allowing text to confirm their assumptions.
Moreover, in extreme cases where text is unavailable, it could facilitate them to grab the general idea without reading the text.
For example, elderly people and people with visual impairment may find it difficult to read the text in the legend clearly.
Language barriers may also prevent users from understanding information from multiple views in cross-language environments.


\smallskip
\noindent
\textbf{Limitation and Future Work.}
First, we address limitations related to our method's scope and implementation. 
Our current focus on categorical and sequential colormaps overlooks other variants like diverging colormaps and excludes considerations of aesthetics, semantics, and mark type.
\re{Similarly, we do not model subjective user preferences, such as user age and cultural background\cite{morita2024color}.
Our optimization-based approach, while effective at balancing complex trade-offs, also lacks the inherent explainability of rule-based systems like KG4Vis\cite{li2021kg4vis}, it can generate an optimal result but cannot provide human-readable rules explaining why.}
On the implementation side, we identify data relationships through specific encodings, which may not scale well or uncover inter-column relationships. 
Our use of Vega-Lite also covers basic mark types but omits complex marks and glyphs.

Second, our evaluation has its own limitations. 
\re{The user study is limited by the confounding of color saturation, as our emphasis on discriminability inherently produces more saturated colors than the baseline. 
Consequently, we cannot fully disentangle a preference for effectiveness from a potential preference for saturation alone. 
Lastly, although we developed an interactive interface, its usability has not been formally evaluated.}

Despite these limitations, our adaptable framework is designed for extension.
\re{For instance, it could be updated to support users with color vision deficiencies by incorporating CVD-safe metrics}, or expanded to include other criteria like name differences in addition to the established CIEDE2000 metric for color discriminability.
We also plan to develop more sophisticated techniques for automatic relationship detection and incorporate libraries like d3.js to support an expanded range of visualizations.
These efforts will enhance our colormap design approach to be more comprehensive.

\section{Conclusion}
This work enhances consistency in multi-view visualizations using color encoding, which simplifies designs by removing the need for extra visual connectors and reducing viewer cognitive load.
We address two main challenges: maintaining effectiveness within individual views while ensuring overall consistency, and translating these needs into measurable design criteria. 
To solve these, we propose a framework using knowledge-based graphs for MV visualizations to unify components and establish design parameters.  
We provide an algorithm for generating optimal colormaps and an interactive interface for user adjustments. Our approach, validated through case studies and a user study, proves more effective and efficient than current techniques.
Our work represents significant progress in enhancing MV visualization interpretability and efficiency, while substantially reducing the colormap design burden for both professional designers and non-expert users.
\re{Future work will enhance our colormap design approach to be more comprehensive.}

\noindent
{\textbf{Acknowledgment}.} \ 
The authors wish to thank anonymous reviewers for their constructive comments. 
The work was supported in part by National Natural Science Foundation of China (62172398).

\bibliographystyle{eg-alpha-doi} 
\bibliography{template}
\newpage
\appendix

\section{Implementation Details}

\subsection{Pseudocode}

\begin{algorithm}[h]
    \small
    \DontPrintSemicolon
    \caption{Pareto-based Genetic Algorithm for Colormap Optimization}
    \label{alg:ga}
    \KwIn{Population size $N$, Max generations $G$, View specifications $V$, Initial seed palette $S$}
    \KwOut{A set of Pareto-optimal colormap solutions $\mathcal{P}_{final}$}
    
    $P_0 \leftarrow \textsc{InitializePopulation}(N, V, S)$\;
    $t \leftarrow 0$\;
    \While{$t < G$}{
        $C_{set} \leftarrow \emptyset$\;
        \ForEach{individual $m$ in $P_t$}{
            Compute fitness score $(C_{SV}(m), C_{MV}(m))$\;
            $C_{set} \leftarrow C_{set} \cup \{(m, [C_{SV}(m), C_{MV}(m)])\}$\;
        }
        $\mathcal{P}_t \leftarrow \textsc{FindParetoFront}(C_{set})$\;
        $P_{next} \leftarrow \{\,m \mid (m, \cdot) \in \mathcal{P}_t\,\}$ \tcp*{Carry over elite solutions}
        
        \While{$|P_{next}| < N$}{
            $(A, B) \leftarrow \textsc{SelectParents}(\mathcal{P}_t)$ \tcp*{Randomly sample}
            $(O_1, O_2) \leftarrow \textsc{Crossover}(A, B)$\;
            $(O_1, O_2) \leftarrow \textsc{Mutate}(O_1, O_2)$\;
            $P_{next} \leftarrow P_{next} \cup \{O_1, O_2\}$\;
        }
        $P_{t+1} \leftarrow P_{next}$\;
        $t \leftarrow t+1$\;
    }
    $\mathcal{P}_{final} \leftarrow \textsc{PostProcess}(\mathcal{P}_t)$\;
    \Return $\mathcal{P}_{final}$\;
\end{algorithm}
\subsection{Metrics and Parameters}
\label{ssec:appendix_params}

To ensure the reproducibility of our optimization framework, this section provides the implementation details regarding our metric handling, key parameters, and threshold settings.

\paragraph{Normalization of Metrics.}
Our cost function consists of four components. To balance their magnitudes, we employ Min-Max Scaling to normalize each cost term to the range of $[0, 1]$. The normalization is calculated as follows:
\begin{equation}
    \small
    \text{norm\_cost} = \frac{\text{cost} - \text{min\_cost}}{\text{max\_cost} - \text{min\_cost}}
\end{equation}
Here, \texttt{cost} is the raw value, while \texttt{min\_cost} and \texttt{max\_cost} are the historical minimum and maximum values recorded across multiple runs. These historical extrema are stored in a \texttt{params.json} file and are independently maintained and updated for each test case. This dynamic updating mechanism allows the normalization to adapt to the cost distributions of different optimization tasks.

\paragraph{Weight Settings.}
The relative importance of the four cost terms can be adjusted by the user. By default, all weights are set to 1, corresponding to single-view color discriminability, color continuity, multi-view color discriminability, and hue uniformity, respectively. This initial setting treats each design criterion with equal importance.

\paragraph{Cost Function Thresholds and Penalty Factor.}
In the calculation of raw costs, we define specific thresholds to distinguish between acceptable and suboptimal color differences. When a calculated value fails to meet a threshold, we apply a smooth penalty using a Sigmoid function.
\begin{itemize}
    \item \textbf{Penalty Factor:} A global \texttt{PENALTY\_FACTOR} is set to \texttt{0.2} to adjust the steepness of the Sigmoid function.
    \item \textbf{Minimum Thresholds:} The predefined thresholds for the cost functions are as follows:
    \begin{itemize}
        \item \texttt{color\_difference\_sv}: 30
        \item \texttt{color\_continuity}: 30
        \item \texttt{color\_difference\_mv}: 20
        \item \texttt{hue\_uniformity}: 20
    \end{itemize}
    \item \textbf{Fitness Evaluation Thresholds:} During the genetic algorithm's fitness evaluation, a secondary check is applied to the \textit{normalized} cost values. The \texttt{max\_allowed\_values} for the four normalized cost terms are set to \texttt{[0.2, 0.2, 0.2, 0.2]}. If a solution's cost exceeds this value, an additional penalty is applied to guide the search toward solutions that perform well across all metrics.
\end{itemize}

\subsubsection{Genetic Algorithm Parameters}
Our genetic algorithm is configured with the following parameters. 
These are based on empirical observations.
\begin{itemize}
    \item \textbf{Population Size:} \texttt{pop\_size = 50}.
    \item \textbf{Number of Generations:} \texttt{generations = 100}.
    \item \textbf{Elite Pool Size:} \texttt{n\_best = 10}. This parameter specifies the number of top-performing individuals selected from the Pareto front to serve as the primary pool for creating the next generation.
    \item \textbf{Selection Mechanism:} We employ an elitist selection strategy. In each generation, all non-dominated solutions are identified to form the Pareto front. From this front, we select up to \texttt{n\_best} individuals to form the elite pool. Parents for reproduction are then randomly sampled from this elite pool.
    \item \textbf{Crossover Rate:} \texttt{0.5}. For any two parents, their color schemes for views at hierarchy level 1 are swapped with a 50\% probability.
    \item \textbf{Mutation Step:} Mutation is performed by applying a small random perturbation to a color's HSV values. The magnitude of this perturbation is controlled by the \texttt{step} parameter, which defaults to \texttt{0.05} and is user-adjustable.
\end{itemize}

\section{Quantitative Evaluation Metrics}

 \label{app:metrics} 
 This appendix provides the mathematical definitions for the three quantitative metrics used in our evaluation (Sect. 5.2). For all metrics, a higher score indicates better performance. 
 
 \paragraph{Worst-Case Discriminability (WCD)} WCD assesses single-view clarity by measuring the minimum perceptual distance between any two colors within a single view's colormap. It quantifies the "weakest link" in color differentiability, where a higher score means even the most similar colors are easy to distinguish. 
 \begin{equation} 
    \small
    \text{WCD}(V) = \min_{c_i, c_j \in C(V), i \neq j} \Delta E_{00}(c_i, c_j) 
\end{equation} Here, $C(V)$ is the set of colors in the colormap for view $V$. For an entire multi-view visualization, the overall WCD is the minimum WCD score across all views. 
 
 \paragraph{Parallel Relationship Score (PRS)} PRS evaluates multi-view consistency for non-hierarchical (i.e., partial-redundancy) relationships. 
 It measures the minimum perceptual distance between any two colors from different views that represent different data entities. 
 A high PRS ensures that unrelated items across the visualization are not perceptually confused with one another. \begin{equation} \small \text{PRS}(\mathcal{V}) = \min_{\substack{c_a \in C(V_i), c_b \in C(V_j) \\ V_i, V_j \in \mathcal{V}, i \neq j \\ \text{key}(c_a) \neq \text{key}(c_b)}} \Delta E_{00}(c_a, c_b) 
\end{equation} 
 Here, $\mathcal{V}$ is the set of all views in the visualization, and $\text{key}(c)$ refers to the underlying data entity that the color $c$ encodes. 
 
 \paragraph{Hierarchical Quality Score (HQS)} HQS assesses the quality of hierarchical colormaps by balancing two competing factors: the internal discriminability of the child colors and their hue similarity to the parent color. It is defined as the ratio of the child colors' worst-case discriminability (Child-WCD) to their average hue deviation from the parent (HHD). 
 \begin{equation} \small \text{HQS} = \frac{\text{Child-WCD}}{1 + \text{HHD}} \end{equation}
 
 The two components are defined as: \begin{itemize} \item \textbf{Child-WCD}: The minimum perceptual distance between any pair of colors in the child set, calculated identically to WCD. \[ \small \text{Child-WCD} = \min_{c_i, c_j \in C_{child}, i \neq j} \Delta E_{00}(c_i, c_j) \] \item \textbf{HHD (Hierarchical Hue Deviation)}: The average absolute difference in hue between the parent color $c_p$ and each child color $c_k$, accounting for the circular nature of the hue space. \[ \small \text{HHD} = \frac{1}{|C_{child}|} \sum_{c_k \in C_{child}} \min(|H(c_p) - H(c_k)|, 360^\circ - |H(c_p) - H(c_k)|) \] \end{itemize} Here, $c_p$ is the parent color, $C_{child}$ is the set of child colors, and $H(c)$ is the hue of color $c$ in degrees.

\section{Extra Cases}

\begin{figure}[htb]
    \centering
    \includegraphics[width=0.95\linewidth]{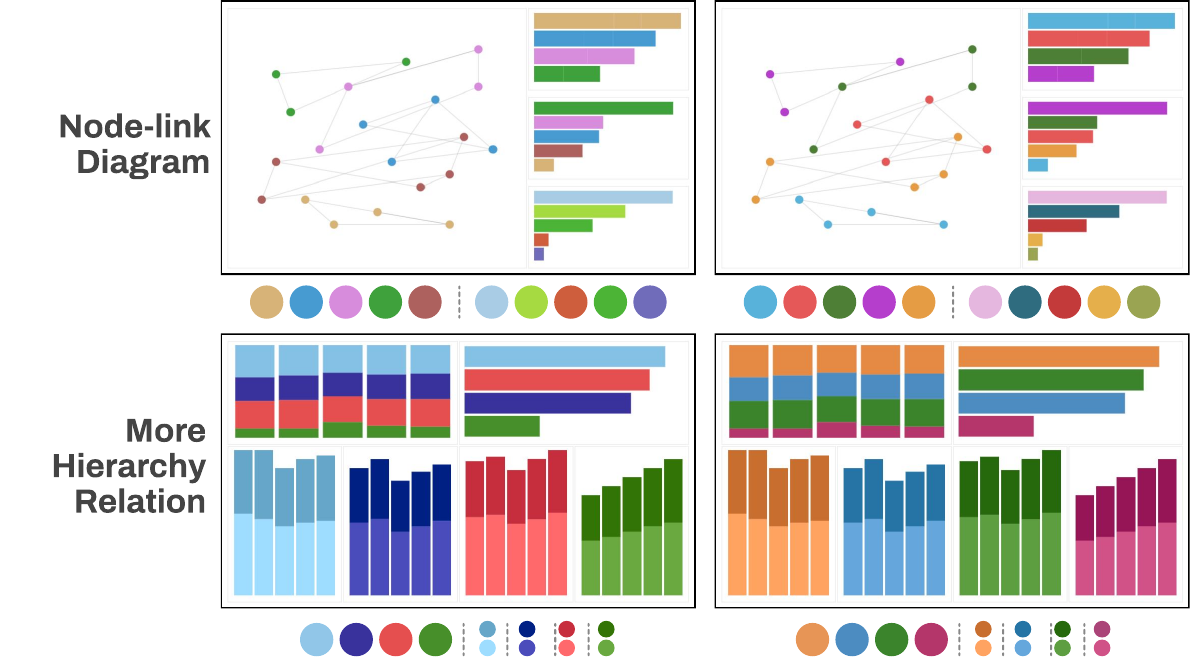}
    \caption{Extra cases including a node-link diagram and a 4 categorical hierarchical case.}
    \label{fig:extracases}
\end{figure}

Our framework is based on modeling abstract data relationships inferred from Vega-Lite specifications. This allows our approach to generalize to more supported chart types.
For instance, the top case in Figure above demonstrates a successful application to a node-link diagram, maintaining color consistency with a corresponding bar chart.
Additionally, we present a solution for a complex hierarchical scenario with four distinct parent categories. This case is inspired by the real-world example of poor hierarchical coloring discussed in our Introduction (Figure 1b).
\end{document}